\documentclass[journal,comsoc]{IEEEtran}
\usepackage{subfigure}
\usepackage{graphicx}
\usepackage{array}
\usepackage{graphicx}
\usepackage{amsmath}
\usepackage{multirow}
\usepackage{ragged2e}
\usepackage{subfig}
\usepackage{xcolor}

\usepackage{algorithm}  
\usepackage{algorithmicx}  
\usepackage{algpseudocode}

\usepackage[T1]{fontenc}

\ifCLASSINFOpdf
\else
\fi
\usepackage{amsmath}
\usepackage[cmintegrals]{newtxmath}
\begin{document}
%
\title{Graph based Nearest Neighbor Search:\\ Promises and Failures}
%
%
%

\author{{Peng-Cheng Lin,
        Wan-Lei Zhao}
\thanks{Peng-Cheng Lin and Wan-Lei Zhao are with the Department
of Computer Science, Xiamen University, Xiamen, China. Wan-Lei Zhao is the corresponding author.
E-mail: wlzhao@xmu.edu.cn}
\thanks{Manuscript received Month Day, Year; revised Month Day, Year.}}

%
%

\markboth{IEEE Multimedia,~Vol.~x, No.~x, Month~Day}%
{Shell \MakeLowercase{\textit{et al.}}: Bare Demo of IEEEtran.cls for IEEE Communications Society Journals}
%



\maketitle

\begin{abstract}
Recently, graph based nearest neighbor search gets more and more popular on large-scale retrieval tasks. The attractiveness of this type of approaches lies in its superior performance over most of the known nearest neighbor search approaches as well as its genericness to various metrics. In this paper, the role of two strategies, namely hierarchical structure and graph diversification that are adopted as the key steps in the graph based approaches, is investigated. We find the hierarchical structure could not achieve ``much better logarithmic complexity scaling'' as it was claimed in the original paper, particularly on high dimensional cases. Moreover, we find that similar high search speed efficiency as the one with hierarchical structure could be achieved with the support of flat \textit{k}-NN graph after graph diversification. Finally, we point out the difficulty, that is faced by most of the graph based search approaches, is directly linked to ``curse of dimensionality''.
\end{abstract}


%
\IEEEpeerreviewmaketitle

\section{Introduction}
\IEEEPARstart{N}{earest} neighbor (NN) search is a task that searches for the closest neighbors from a group of given candidates for a query. It plays fundamental role in various applications, such as multimedia information retrieval, data mining, computer vision and machine learning. The early exploration about this issue could be traced back to 1970s~\cite{kdtree75}. However, it is still a new issue in the sense that it becomes more and more imminent recently given the big data problem arises particularly from various multimedia applications. This leads to the timely demand in scalable and efficient nearest neighbor search approach on large-scale and high dimensional data. This decades-old issue is therefore shed with new light. However, due to the ``curse of dimensionality'', effective solution is still slow to occur.

In the nearest neighbor search, both the query and the candidate samples are assumed to be from the same space \textit{i.e.}, $R^d$. The closeness between samples is usually predefined by a metric $m(\cdot,\cdot)$. The time complexity of brute-force search is $O(n{\cdot}d)$, where \textit{n} is the number of candidates. Although it is linear with respect to either \textit{n} or \textit{d}. Its computational complexity is very high when both \textit{n} and \textit{d} are large. Research on this issue aims to explore approach whose time complexity is much lower than $O(n{\cdot}d)$.

This issue has been traditionally addressed by a variety of tree partitioning approaches, such as K-D tree~\cite{kdtree75}, and R-tree~\cite{rtree84}, etc. Although they are different in details, they are all designed to partition the space into hierarchical sub-spaces. The nearest neighbor search only traverses over a few branches of the sub-spaces. Unfortunately, unlike B-tree in 1D case, the partition scheme does not exclude the possibility that nearest neighbor resides outside of these candidate sub-spaces. Therefore, extensive probing over large number of branches in the tree becomes inevitable. Recent indexing structures FLANN~\cite{pami14:flann} and Annoy~\cite{cvpr08:annoy} partition the space with hierarchical \textit{k}-means and multiple K-D trees respectively. Although they are efficient, sub-optimal results are achieved.

Apart from tree partitioning approaches, quantization based approaches~\cite{JDS11,artem16} and locality sensitive hashing (LSH)~\cite{lsh04} have been extensively explored in the last decade. Approaches from both categories save up lot of memory consumptions and are very efficient when the size of encoding bits is short. Nevertheless, the memory efficiency and speed-up achieved by approaches from both categories are very limited when the search accuracy is required to be high. Another major disadvantage for these approaches is that they are mostly only suitable for $l_p$-norms. Furthermore, the design of generic hash functions is non-trivial.

Recently, graph based approaches such as nearest neighbor descent (NN-Descent)~\cite{weidong} or hill-climbing (HC)~\cite{icai11:kiana}, demonstrate superior performance over other categories of approaches in many large-scale NN search tasks~\cite{pami18:yury, dpg:wenli}. All the approaches in this category are built upon an approximate \textit{k}-NN graph. The search procedure starts from a group of random seeds and traverses iteratively over the graph by the best-first search. Guided by the neighbors of visited vertex's, the search procedure descents closer to the true nearest neighbor in each round until no better candidates could be found. Approaches in~\cite{icai11:kiana,pami18:yury,infosys13:yury} in general follow the similar search procedure. The major difference between them lies in the structure of \textit{k}-NN graphs upon which the NN search is undertaken. For most of the graph based approaches, the extra merit is that they are suitable for various distance measures.

On one hand, graph-based approaches get more and more popular recently. On the other hand, the overall behavior about them is yet unclear. In this paper, a comprehensive experimental study about graph based approaches is presented. The most popular NN search approaches in recent years, namely hierarchical navigable small world (HNSW) graphs~\cite{pami18:yury} and diversified proximity graph (DPG)~\cite{dpg:wenli} are taken as the study cases. Firstly, we confirm that the graph based approaches are apparently superior than approaches of other categories on large-scale datasets of various types. Moreover, we show where the graph based NN search works and where it fails. Namely, we study the role of hierarchical structure and graph diversification in the graph based approaches. Contradictory to the claim made in \cite{pami18:yury}, we find the hierarchical structure is not helpful for the nearest neighbor search when the data dimension goes high. Similar or even better performance could be achieved by diversified flat graph. Although graph-based approaches are effective in many scenarios, we further point out that all the graph-based approaches face the same difficulty as the data dimension is high. This difficulty is directly linked to the ``curse of dimensionality''.

\section{Related Work}
There is a significant literature on the nearest neighbor search algorithms as the result of more than four decades of continuous research. They in general fall into four categories. Namely, they are tree-based space partition approaches, quantization-based approaches, locality sensitive hashing approaches and graph-based approaches. In this section and the next, the general idea as well as the advantages and pitfalls of approaches in each category are reviewed.

\subsection{Tree-based Space Partition Approaches}
The basic idea of this category is to partition the space into embedded hierarchical sub-spaces. This search process is conducted efficiently by using the tree properties to eliminate large portions of the branches. One commonly used method, the K-D tree~\cite{kdtree75} is very efficient in low dimensional case. However, as the dimension increases, the performance of K-D tree degrades quickly since the query process has to visit nearly every nodes in the tree. Vantage point tree (VP-tree) \cite{vldb15:extendvp} is proposed to enhance its performance by introducing a better partition strategy. FLANN~\cite{pami14:flann} and Annoy~\cite{annoy} are two representative space partition algorithm proposed in recent years. FLANN partitions the space via hierarchical \textit{k}-means. While Annoy partitions the space by a variant of hierarchical \textit{k}-means. Although they are efficient even on high dimensional data, sub-optimal results are achieved. In our comparative study, both FLANN and Annoy are selected as the representative tree-based approaches.

\subsection{Quantization-based Approaches}
Quantization-based search approaches have been extensively explored in recent years~\cite{JDS11,artem16}. In general, the candidate vectors are all compressed via vector (or sub-vector) quantization. This makes it possible to hold the whole large-scale candidate dataset in the memory. The NN search is conducted between the query and the quantized candidate vectors. The distance between query and candidates is approximated by the distance between query and vocabulary words that are used for quantization. Due to the considerable compression loss on the candidate vectors, high search quality is hardly desirable. In our study, product quantization~\cite{JDS11} is selected as the representative approach of this category.

\subsection{LSH-based Approaches}
Approaches of this category are derived from the same idea of locality sensitive hashing~\cite{lsh04}. Multiple hash tables are predefined to map vector into a series of hash codes. The hash codes of the query sample are compared to the hash codes of all the candidate vectors. Vectors closer to a query are expected to share the same or similar hash codes as the query. These closer candidates are kept to compare with the query vector directly. In this sense, the hashing acts as a filtering scheme for efficient computation.

In general, more number of hash tables leads to preciser hash encoding, while inducing higher computation and memory consumption. In addition, the performance is highly dependent on the design of hash functions. In the early literatures \cite{TSCG04:Locality}, the efforts are devoted to propose various LSH families for different types of metrics. Later, learning-based hashing approaches~\cite{cvpr12:SupHash} take the data distribution into account. Most of these approaches require large amount of space for indexing that limits the scalability of the approach. Recently, a projection-based approach named SRS~\cite{srs14} is proposed. It requires only a single tiny index. SRS shows high scalability and is the state-of-the-art in this category of techniques. Unfortunately, it is only suitable for $l_{2}$-space. In our study, SRS is selected as the representative approach of this category.

\subsection{Graph-based Approaches}
Recently, graph-based approaches~\cite{weidong,pami18:yury,infosys13:yury,dpg:wenli} have drawn considerable attention. Among all these approaches, the key step lies in the construction of the graph structure.
A candidate vector is represented as vertex in the graph. Two vertices are connected by an edge when they are sufficiently close from one to another. Usually several tens of neighbors are kept for each vertex. Once the graph is constructed, most of the approaches follow similar search procedure~\cite{icai11:kiana}. It starts from one or several random seeds (vertices) and explores the neighborhoods of visited vertices. The search descents to closer samples if any closer samples are discovered. The process repeats until no closer samples are found. As demonstrated in different contexts, graph-based approaches~\cite{weidong,pami18:yury,dpg:wenli,infosys13:yury} show considerably superior performance over approaches of other categories. 

Since the focus of this paper is to study the behavior of graph-based approaches. A more detailed review on the representative approaches in this category is presented in the next section.

\section{Review on Graph-based Nearest Neighbor Search}
\label{sec:hnsw}
As pointed out in the previous section, most of the graph-based approaches follow similar search procedure. The major difference between them lies in the construction of graph, upon which the nearest neighbor search is undertaken. The most representative graph construction approaches are KGraph~\cite{weidong} and HNSW~\cite{pami18:yury}, both of which are generic to various distance measures.

\textbf{\textit{k}-nearest neighbor Graph (KGraph)}~\cite{weidong}. The idea of KGraph is inspired by the observation that ``a neighbor of a neighbor is also likely to be a neighbor''. It starts by generating \textit{k} random neighbors for each sample. The construction process iteratively refines this raw \textit{k}-nearest neighbor graph by performing pair-wise comparison within each \textit{k}-NN and reverse \textit{k}-NN list. The process terminates when no improvement could be made. In the search step, the hill-climbing search strategy~\cite{icai11:kiana} is adopted.

During the hill-climbing search, samples in the visited \textit{k}-NN list will guide the search to closer neighborhood of the query. As observed in~\cite{dpg:wenli} and~\cite{cvpr16:ben}, if samples in the list are close to each other, it will guide the search to reach to the same location. As a result, it is redundant to compare the query to all the neighbors of a visited vertex. Diversified proximity graph (DPG)~\cite{dpg:wenli} is proposed recently to enhance the sparsity of the neighborhood. Basically, it removes the samples from the neighborhood of one vertex that are closer to other samples in the neighborhood than they are to the vertex. As shown in~\cite{dpg:wenli}, DPG boosts the performance of NN search considerably.

\begin{figure}
	\begin{center}
		\includegraphics[width=0.8\linewidth]{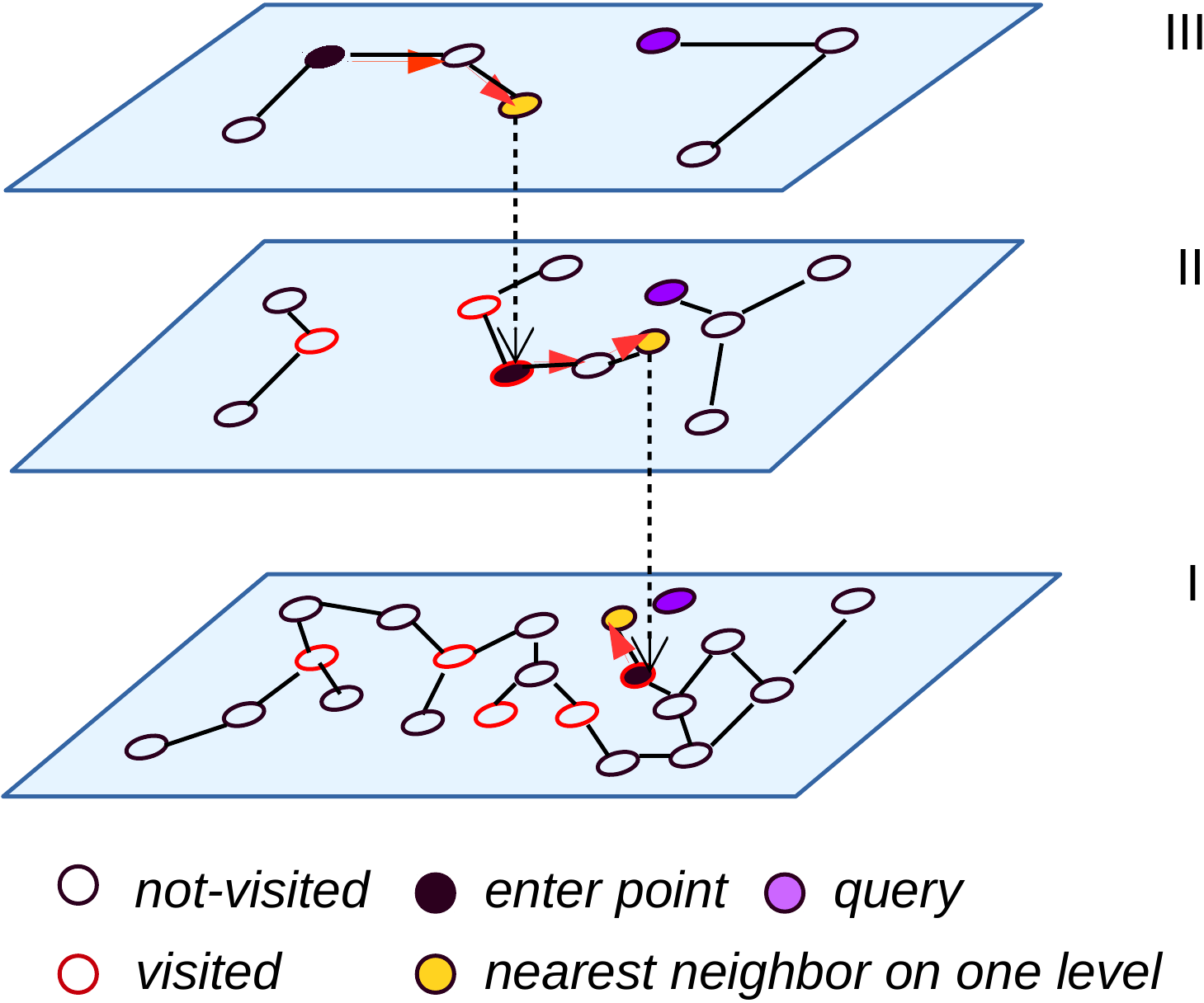}
	\end{center}
	\caption{The illustration of hierarchical structure of HNSW (three layers case). Each layer of HNSW is an NSW graph. The long-range links are maintained on top layers. The search starts from the top layer. The found nearest neighbor from non-bottom layers is treated as the enter point of the NN search on the lower layer.}
\label{fig:hnsw}
\end{figure}

\textbf{Hierarchical navigable small world (HNSW) graphs}~\cite{pami18:yury}.
The basic idea of HNSW is inspired from the small world graph theory. The graph keeps both long range links\footnote{The terms ``link'', ``connection'' and ``edge'' are inter-changeable across this paper.} to far neighbors and short range links to close neighbors. The links are organized into an hierarchy. The long range links are maintained on the top layer while the short range links are maintained on the lower. The search is fulfilled from the top layer to the lower. According to the paper~\cite{pami18:yury}, such kind of coarse-to-fine search leads to ``a much better logarithmic complexity scaling'' than that of based on flat graph. Since the codes were released on GitHub\footnote{https://github.com/nmslib/hnswlib}, HNSW already receives more than \textit{300} stars in two years. The search approach so far has been integrated with various search systems.

The hierarchical structure of HNSW is illustrated in Fig.~\ref{fig:hnsw}. As seen from the figure, the search starts from a fixed sample on the top layer. The NN search explores the \textit{k}-NN list of a visited vertex, and descents greedily to a closer neighbor if there exists. The search on one layer stops as no closer neighbor could be found. The discovered closest neighbor on the current layer is treated as the starting point (referred as ``enter point'' in~\cite{pami18:yury}) of the search on the lower layer. Such kind of top-down greedy search continues until it reaches to the bottom layer. On the bottom layer, the search takes the closest neighbor found from upper layer as the starting point. The search moves towards the query each time by expanding neighbors of vertices' in a maintained top-\textit{k} list. The top-\textit{k} list is updated as long as any closer top-\textit{k} neighbor is found. The recall of the search procedure is controlled by the parameter \textit{k} (referred as \textit{ef} in~\cite{pami18:yury}).


The construction of HNSW graphs follows the same procedure as the NN search over it. Each candidate sample is treated as a query to query against the HNSW graphs under construction. HNSW graphs are incrementally built by repetitively inserting candidate samples into the hierarchy. The probability that a query is inserted to a layer is regularized by an exponentially decaying probability distribution. The lower the layer is, the higher of the chance that the query is inserted. On the layer that a query sample should be inserted, the discovered \textit{M} neighbors are kept in the neighbors list of this sample. Accordingly, this query is possibly inserted into the neighbors lists of these \textit{M} neighbors'. Parameter \textit{M} controls the scale of the neighborhood list. 

\begin{figure}
	\begin{center}
		\includegraphics[width=0.35\linewidth]{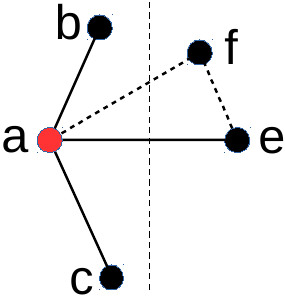}
	\end{center}
	\caption{The illustration of the heuristic strategy used to select the graph neighbors in the construction of HNSW graphs. An edge from $a$ to $e$ occludes an edge from $a$ to $f$ since $f$ is closer to $e$ than $a$.}
	\label{fig:delaunay}
\end{figure}

\textbf{Graph diversification} (GD) During the the insertion of a sample into the graph of each layer,  the insertion operation is verified by a heuristic pruning strategy to enhance the diversity of a vertex's neighborhood. This pruning strategy is shown in Fig.~\ref{fig:delaunay}. As seen in the figure, the neighborhood of sample \textit{a} are already inserted with samples \textit{b} and \textit{c}. For an incoming sample \textit{e}, we check whether it should be inserted into the \textit{k}-NN list of \textit{a}. According to the strategy, sample \textit{e} is allowed to join the neighborhood \textit{iff} $m(e,a) < m(e, b)$ and $m(e,a) < m(e, c)$. Similarly given \textit{e} is inserted, if another sample \textit{f} is going to join the graph, its distance to \textit{a} should be smaller than the distances of all joined samples (samples \textit{b}, \textit{c} and \textit{e}) to sample \textit{f}. It is imaginable that HNSW graph of each layer has been diversified by this strategy.

The motivation behind this strategy is to avoid comparing query to the cluttered neighbors, which is similar as DPG~\cite{dpg:wenli}. If samples are close to each other, it is redundant to visit them all since they offer similar amount of guiding information for a query. From now on, this strategy is called as graph diversification (GD). This strategy is also adopted in~\cite{cvpr16:ben} for graph diversification. 

In this paper, a hybrid scheme is presented. Namely, GD is applied on flat approximate \textit{k}-NN graph. In the experiment section, we are going to show superior search performance is achieved when the NN search is conducted with the support of GD diversified flat \textit{k}-NN graph.

\section{Overview of Comparative Study}
\label{sec:over}
In our first study, graph based NN search approaches, namely HNSW~\cite{pami18:yury}, KGraph~\cite{weidong} and DPG~\cite{dpg:wenli} are compared to non-graph based approaches. Secondly, we investigate the role of hierarchical structure in HNSW. As claimed in~\cite{pami18:yury}, the hierarchy helps to skip many far neighbors in the early search. The hierarchy seemingly achieves ``logarithmic complexity scaling'' as B-tree does on one dimensional data. In our study, the search performance of HNSW is compared to the one based on flat graph, which is exactly the bottom layer of HNSW graphs. The hill-climbing search starts from \textit{k} (referred as \textit{ef} in~\cite{pami18:yury}) number of random samples for both cases.

Additionally, we investigate the effectiveness of graph diversification, namely GD and DPG~\cite{dpg:wenli}. The performance of HNSW is compared with that of \textit{k}-NN graph after GD and \textit{k}-NN graph after DPG respectively. For GD graph, it is produced in three steps. Firstly, approximate \textit{k}-NN graph \textit{G} is built for candidate data by package KGraph provided by~\cite{weidong}. Graph \textit{G} is further diversified by the pruning strategy introduced in HNSW (illustrated in Fig.~\ref{fig:delaunay}). For a graph with \textit{L} approximate neighbors,  $L/2$ neighbors at most are selected for each sample. For diversified graph $G^*$, the reverse neighbors of each sample in $G^*$ are appended to the diversified neighborhood. Since overlapping happens between the neighbors and reverse neighbors, we actually take the union of diversified neighbors and reverse neighbors. For DPG graph, we simply follow the way presented in~\cite{dpg:wenli}. For GD and DPG, they are built upon the same \textit{k}-NN graph produced by KGraph~\cite{weidong}.

As one will see, based on the above two experiments, the claimed ``much better logarithmic complexity scaling'' with hierarchical structure~\cite{pami18:yury} is not observable for HNSW on high dimensional data. Another experiment is conducted to figure out the reason why HNSW and other graph based approaches fail on high dimensional case. 
  
\begin{table*}
\begin{center}
\caption{Overview of Datasets}
\footnotesize{
\begin{tabular}{|l|c|c|c|c|c|c|}
\hline
Dataset & n &d & Qry & Exh. Srch (s) & LID~\cite{nips05:Levina} & Type \\
\hline \hline
RAND10M4D & $1\times10^7$ & 4 &1,000&59.3& 3.6&synthetic   \\ \hline
RAND10M8D & $1\times10^7$ & 8 &1,000&75.6& 6.5&synthetic   \\ \hline
RAND10M16D & $1\times10^7$ & 16 &1,000&91.8&11.6 &synthetic   \\ \hline
RAND10M32D & $1\times10^7$ & 32 &1,000&147.7&19.4 &synthetic   \\ \hline
RAND1M & $1\times10^6$ & 100 &1,000&34.0&48.9 &synthetic  \\ \hline \hline

SIFT1M~\cite{JDS11} & $1\times10^6$   & 128 & 10,000 &390.1& 16.3&SIFT \\ \hline
GIST1M~\cite{JDS11}& $1\times10^6$ & 960 &1,000&283.1&38.1 &GIST \\ \hline
GloVe1M~\cite{glove14} & $1.2\times10^6$ & 100 &1,000&37.2 & 39.5 &Text\\ \hline 

\end{tabular}
}
\label{tab:datasets}
\vspace{-0.2in}
\end{center}
\end{table*}

\section{Experiments}
\subsection{Experiment Setup}
The experiments are conducted on eight datasets. Namely, there are three real-world datasets, SIFT1M, GIST1M and GloVe1M, which are popularly adopted in NN search evaluation~\cite{JDS11,dpg:wenli}. Meanwhile, five synthetic datasets are adopted. The data dimension ranges from \textit{4} to \textit{100}. For the synthetic data, each data dimension is randomly sampled from the range [0, 1). Datasets RAND1M, GIST1M and GloVe1M are marked as challenging datasets in~\cite{dpg:wenli}. The search performance is measured by \textit{Recall@1} across all our experiments. The brief information about all eight datasets (both queries and candidate set) are summarized in Tab.~\ref{tab:datasets}.  
In addition, the time costs for exhaustive NN search on each dataset are attached. $l_2$-norm is adopted for all synthetic datasets, SIFT1M and GIST1M. \textit{Cosine} distance is adopted for GloVe1M.

In general, the dimension \textit{d} is higher than its real data dimension since data are embedded in a sub-space of \textit{d}-dimensional space~\cite{nips05:Levina,dpg:wenli}. The local intrinsic dimension (LID) is proposed to define the dimension of such sub-space~\cite{nips05:Levina}. In Tab.~\ref{tab:datasets}, LID of each dataset estimated by~\cite{nips05:Levina} is shown in the \textit{6}th column. As shown in the table, LIDs of synthetic data are roughly \textit{1.5}$\sim$\textit{2.0} times lower than their data dimension \textit{d}. This trend is stable. In contrast, LIDs of three real-world datasets are much lower than their data dimension and they vary considerably from one data type to another. As revealed in~\cite{dpg:wenli}, high speed-up is achieved for graph based approaches on dataset whose ratio between \textit{d} and LID is high. While the NN search is challenging when both dimension \textit{d} and LID are high. To this end, we see the advantage of incorporating a series of synthetic datasets in the evaluation. By this way, the performance trend of NN search is more observable with respect to the incremental increase of real data dimension.

The approaches considered here are implemented in C++ and compiled with GCC \textit{5.4}. The experiments are conducted on workstation with CPU E5-2620 \textit{2.4}GHz Xeon and \textit{110}G memory setup. The multi-threads,
SIMD and pre-fetching instructions are disabled in the codes for all the approaches. There are two parameters (namely \textit{M} and \textit{ef}) in HNSW construction. They are optimized according to the suggestions from project \textit{hnswlib}. 

\subsection{Graph-based versus Other Categories}
\label{sec:exp0}

In the first experiment, the performance of graph-based approaches are studied in comparison to NN search of other categories in four datasets. Three of them are marked as ``hard'' level in~\cite{dpg:wenli}. In the comparison, six representative approaches are presented, they are graph-based approaches KGraph~\cite{weidong}, HNSW~\cite{pami18:yury} and DPG~\cite{dpg:wenli}, tree-based approaches FLANN and Annoy, LSH-based approaches SRS and quantization-based approaches PQ. Notice that DPG shares the same approximate \textit{k}-NN graph as KGraph. The graph used in DPG is supplied by KGraph and further diversified. The parameter settings for all these approaches are optimized following the original papers or released package settings. Due to the big performance gap between different category of approaches, the performance is reported as the speed-up times when \textit{Recall@1} reaches to the level of \textit{0.8} and \textit{0.9} respectively.

\begin{figure}
\begin{center}
		 	\subfigure[Recall@1=0.8]
  {\includegraphics[width=0.485\linewidth]{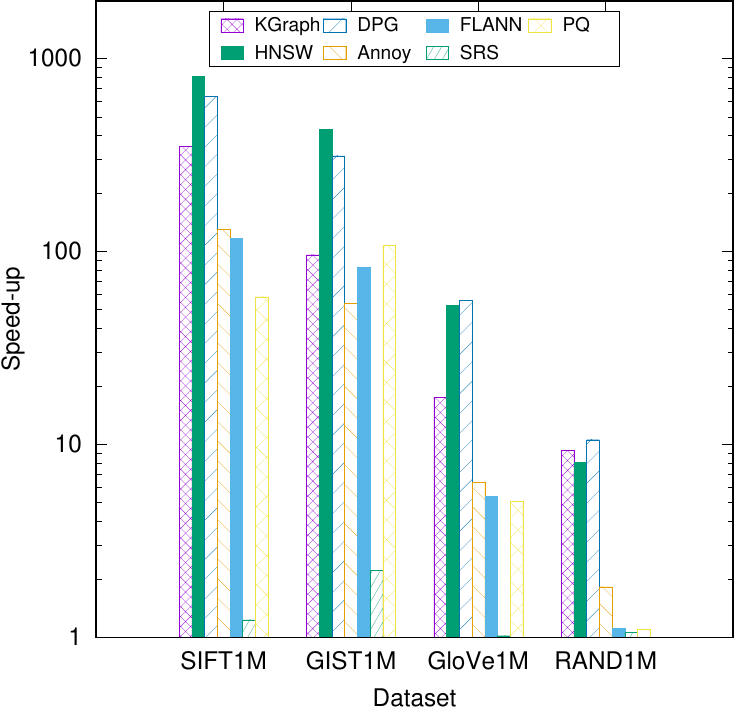}}
  \hspace{0.01in}
	\subfigure[Recall@1=0.9]
  {\includegraphics[width=0.48\linewidth]{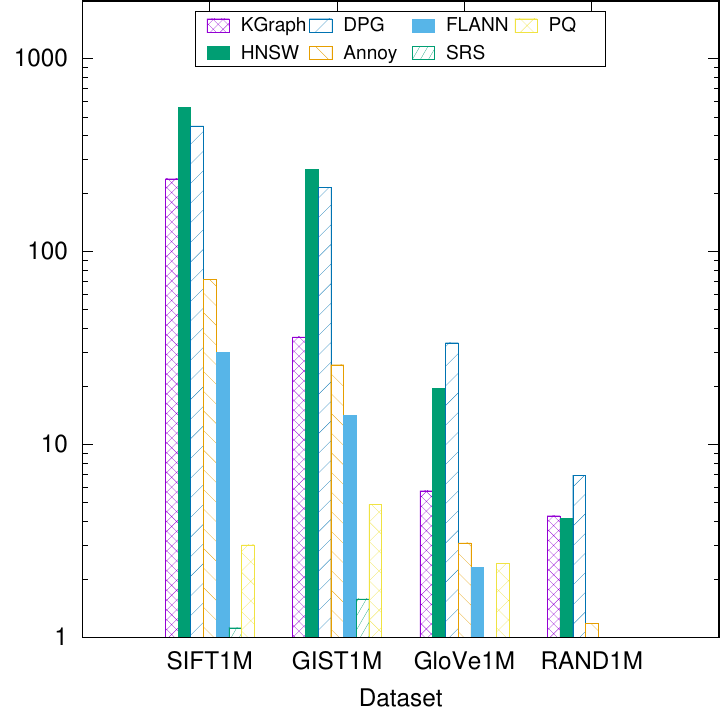}}
	\end{center}
	\caption{Comparison between graph-based approaches and NN search of other categories.
\label{fig:all}}
\end{figure}

As shown in Fig~\ref{fig:all}, KGraph, HNSW and DPG show superior performance over non-graph based approaches across all the datasets. Compared to non-graph based approaches, the latent local structures of samples have been maintained in the graph. The search process is able to proceed quickly by making use of the latent structures.
Among three graph based approaches, HNSW shows the best performance on SIFT1M and GIST1M, while DPG shows slightly better performance on GloVe1M and RAND1M. Another observable trend is that the performance drops steadily for all the considered approaches as the intrisinc dimension (LID) (shown in \textit{6}th column of Tab.~\ref{tab:datasets}) of the data increases.

\subsection{Hierarchy vesus Flat Graph}
\label{sec:exp1}
In the second experiment, we further investigate the performance difference between NN search on HNSW graphs and NN search on flat graph. The flat graph is exactly the bottom layer of HNSW graphs. In the hierarchical search (given as HNSW), the search procedure from ``hnswlib'' package is adopted, which is the same as~\cite{pami18:yury}. In the NN search on flat graph (given as flat-HNSW), NN-Descent procedure from KGraph package~\cite{weidong} is adopted. The major difference from the one used in HNSW lies in the selection of starting points. For HNSW, the starting points are supplied from non-bottom layer search. While for flat-HNSW, the starting points are random seeds.

\begin{figure}
\begin{center}
 \subfigure[RAND10M4D]
  {\includegraphics[width=0.48\linewidth]{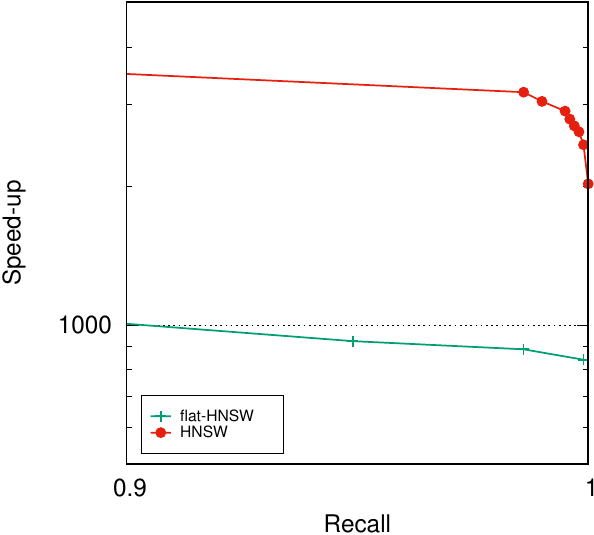}}
  \hspace{0.01in}
	\subfigure[RAND10M8D]
  {\includegraphics[width=0.48\linewidth]{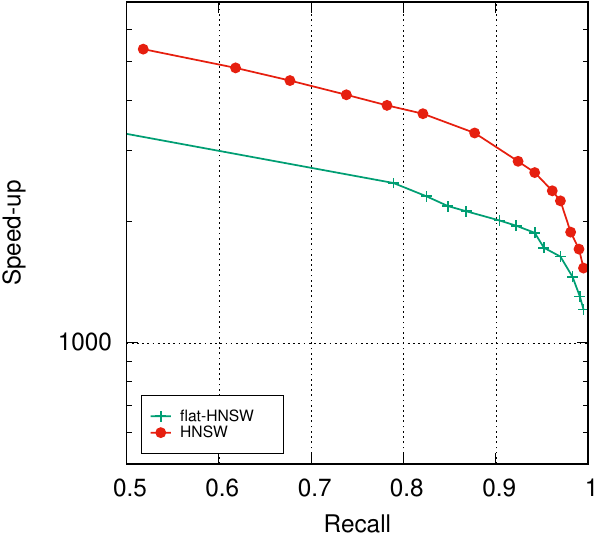}}
  \subfigure[RAND10M16D]
  {\includegraphics[width=0.48\linewidth]{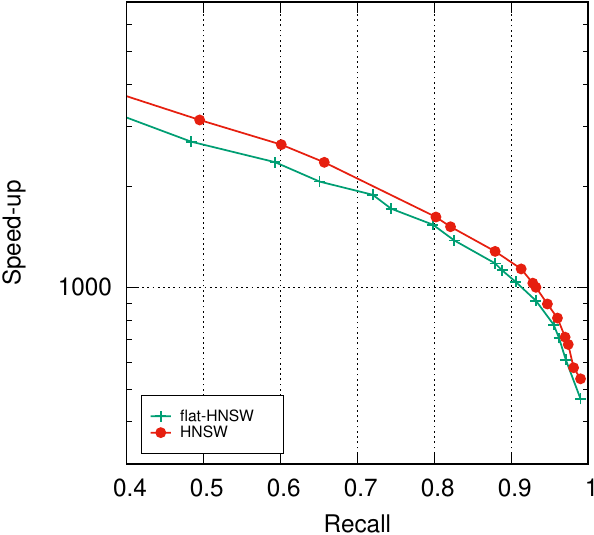}}
  \hspace{0.01in}
	\subfigure[RAND10M32D]
  {\includegraphics[width=0.48\linewidth]{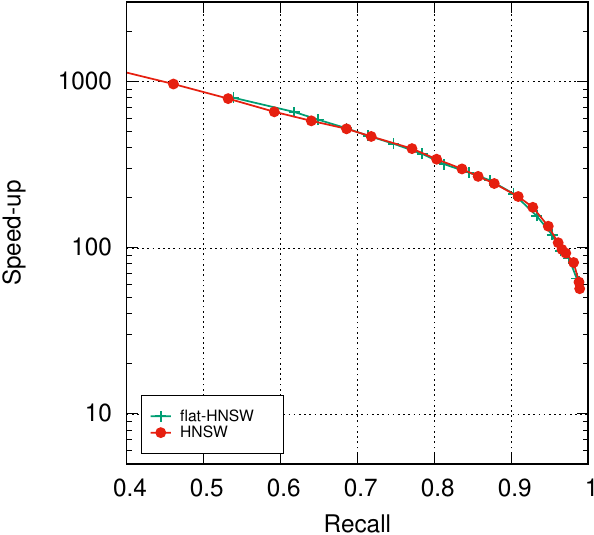}}
  \subfigure[RAND1M]
  {\includegraphics[width=0.48\linewidth]{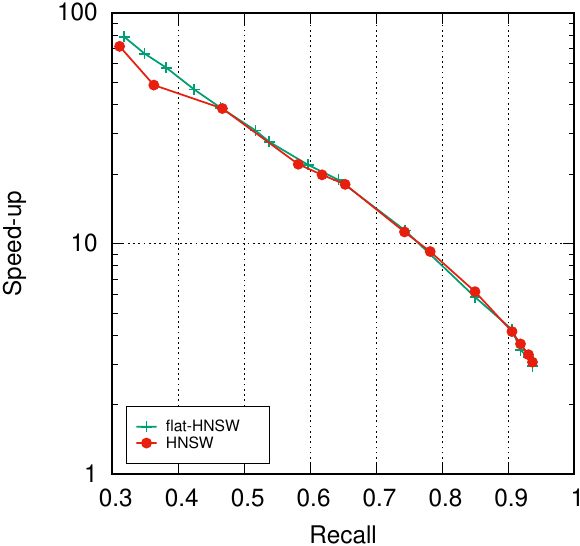}} 
  \hspace{0.01in}  
  \subfigure[SIFT1M]
  {\includegraphics[width=0.48\linewidth]{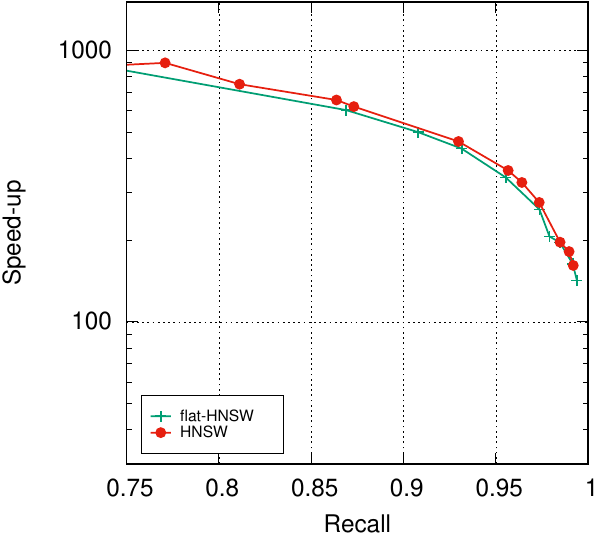}}
  \hspace{0.01in}
	\subfigure[GloVe1M]
  {\includegraphics[width=0.48\linewidth]{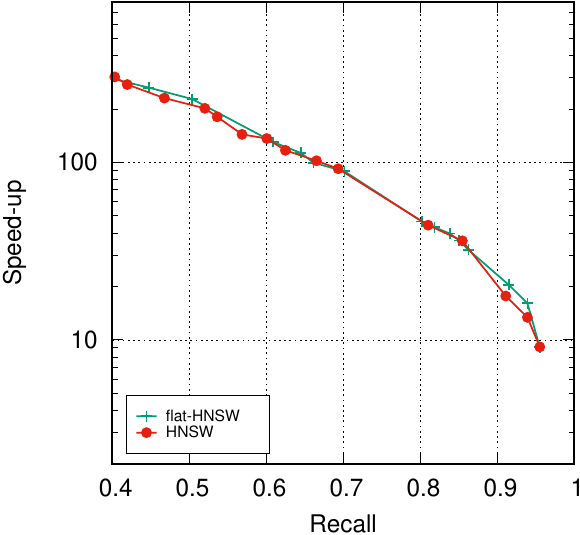}}
  \subfigure[GIST1M]
  {\includegraphics[width=0.48\linewidth]{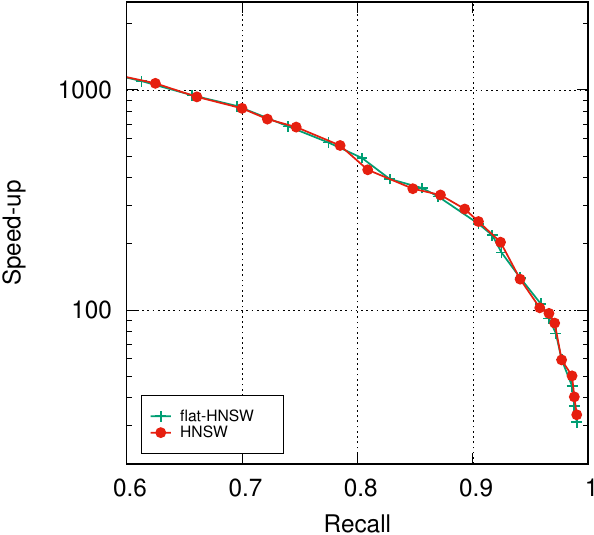}}
  \caption{The NN search performance comparison between HNSW and flat-HNSW on eight datasets.}
\label{fig:hnsw_vs_flat}
\end{center}
\end{figure}

The performance evaluation is presented in Fig.~\ref{fig:hnsw_vs_flat}. The performance is reported as the recall curve against the speed-up achieved over exhaustive search. Fig.~\ref{fig:hnsw_vs_flat}(a)-(e) show the performance of two approaches on synthetic random datasets. While the results on three million-level datasets from real-world are presented on Fig.~\ref{fig:hnsw_vs_flat}(f)-(h).

As shown in the figure, HNSW is around two times faster than flat-HNSW on \textit{4}-dimensional data. While the efficiency of HNSW over flat-HNSW drops significantly as the data dimension reaches to \textit{8}. When the data dimension goes up as high as \textit{32}, there is no considerable difference between these two approaches. A clear trend is observable that the superiority of HNSW fades away as the dimension goes up higher. Specifically, the hierarchical search is no longer helpful when the data is beyond medium dimension (\textit{e.g., 32}). 

On four high-dimensional datasets, the performance of HNSW and flat-HNSW is similar to each other. It is interesting to see the speed-ups on GloVe1M and RAND1M are different while they are on the same dimensionality and the same scale. The speed-up on GloVe1M is considerably higher than that of on RAND1M. This is mainly due to the difference in local intrinsic dimension. LID of GloVe1M is estimated as \textit{39.5} as shown in Tab.~\ref{tab:datasets}. Although this estimation is imprecise, it basically indicates the search on GloVe1M is undertaken on a much lower sub-space. As explained in~\cite{dpg:wenli}, the approximate \textit{k}-NN graph is able to capture the structure of data distribution in each neighborhood of the candidate samples. When the data distributed in the embedded sub-space, such kind of sub-space can be presented as inter-chained \textit{k}-NN lists in the graph. NN-Descent therefore moves along the sub-spaces instead of exploring exhaustively the whole space, which leads to significantly higher efficiency. This is actually the advantage of graph based approaches over other categories of NN search~\cite{pami18:yury,dpg:wenli}. As observed from the results on real-world datasets, both HNSW and flat-HNSW are able to take advantage of the latent data distribution. The hierarchical structure in HNSW brings no extra bonus.

\subsection{Hierarchy versus Graph Diversification}
\label{sec:exp2}
\begin{figure}[t]
\begin{center}
 	\subfigure[RAND10M4D]
  {\includegraphics[width=0.48\linewidth]{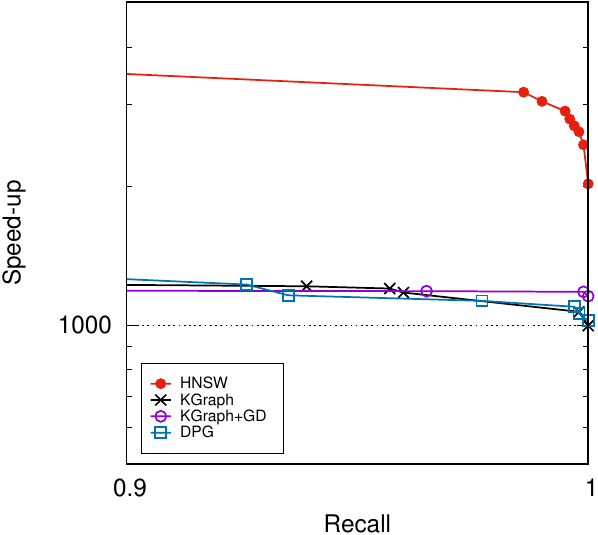}}
  \hspace{0.01in}
	\subfigure[RAND10M8D]
  {\includegraphics[width=0.48\linewidth]{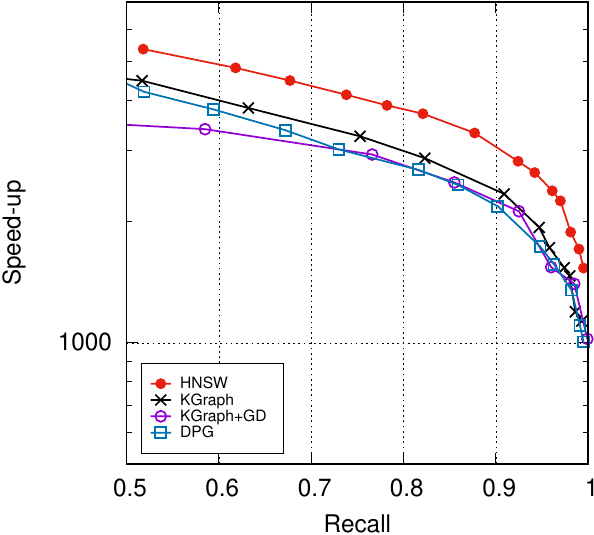}}
  \subfigure[RAND10M16D]
  {\includegraphics[width=0.48\linewidth]{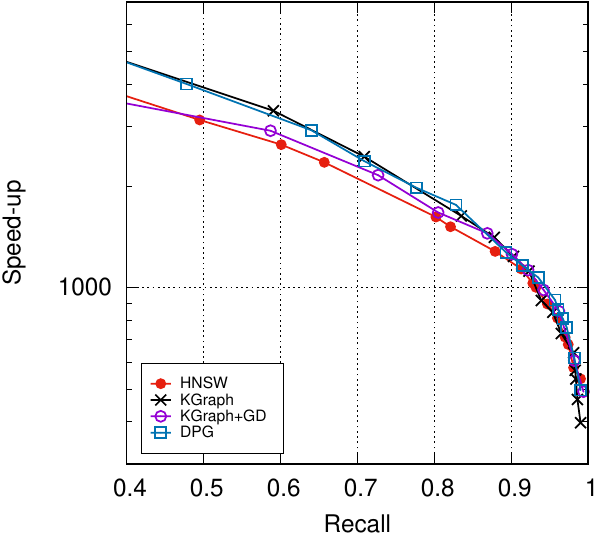}}
  \hspace{0.01in}
	\subfigure[RAND10M32D]
  {\includegraphics[width=0.48\linewidth]{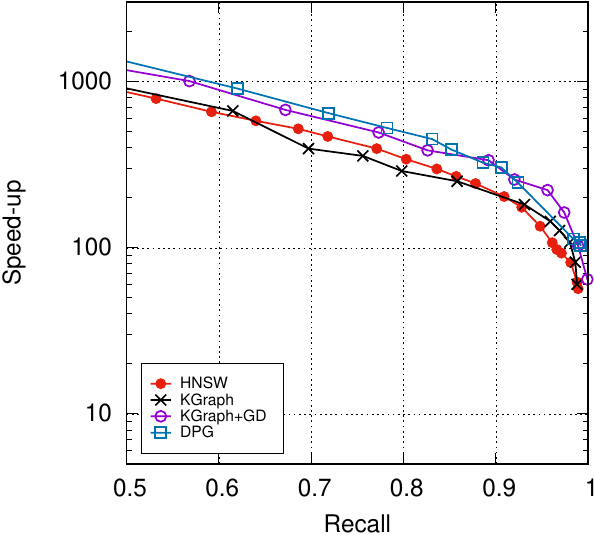}}
	\subfigure[RAND1M]
  {\includegraphics[width=0.48\linewidth]{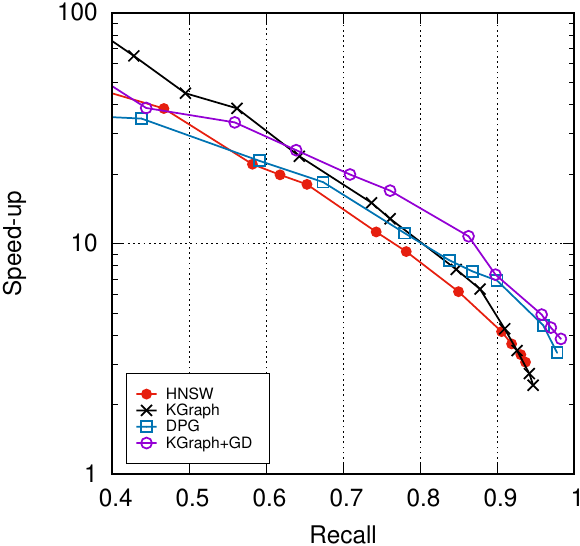}}
  \subfigure[SIFT1M]
  {\includegraphics[width=0.48\linewidth]{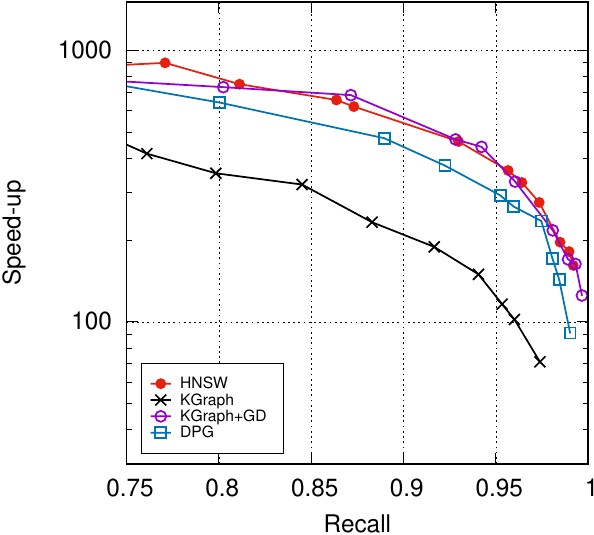}}
  \hspace{0.01in}
	\subfigure[GloVe1M]
  {\includegraphics[width=0.48\linewidth]{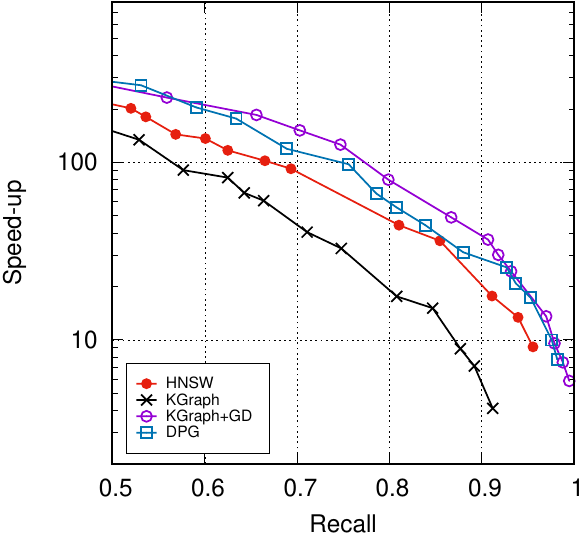}}
  \subfigure[GIST1M]
  {\includegraphics[width=0.48\linewidth]{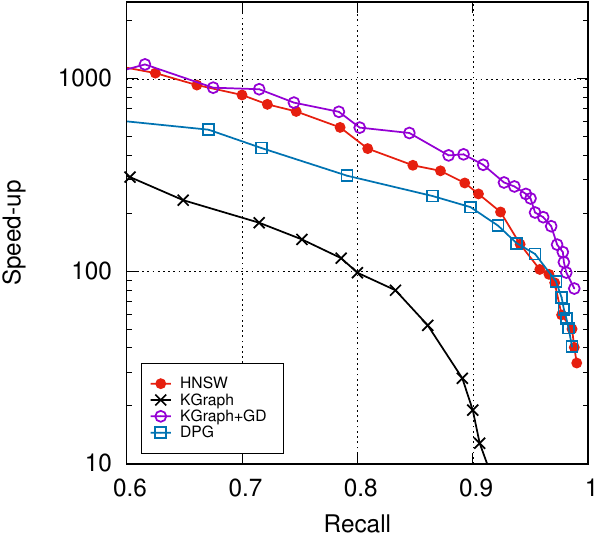}}
  \hspace{0.01in}
  \caption{The NN search performance comparison between HNSW and the one based on approximate \textit{k}-NN graph on eight datasets.}
\label{fig:hnsw_vs_app}
\end{center}
\end{figure}
It is now clear to see hierarchical structure is not helpful for NN search on high dimensional data. In this experiment, we further study the role of graph diversification in the fast NN search. In the experiment, NN search based on HNSW graphs is treated as comparison baseline. Its performance is compared to NN search with the support of flat graph after diversification. Two diversification strategies are considered, namely GD and DPG. On the first hand, KGraph package~\cite{weidong} is called to produce approximate \textit{k}-NN graph. In one configuration, the \textit{k}-NN graph is undergone further graph diversification (GD) that is described in Section~\ref{sec:hnsw}. This configuration is given as ``KGraph+GD''. In another configuration, the \textit{k}-NN graph is diversified by DPG~\cite{dpg:wenli}, which is given as ``DPG'' in the experiment. For reference, the NN search performance with the \textit{k}-NN graph produced by~\cite{weidong} is also presented. It is given as ``KGraph'' in the experiment. So ``KGraph'', ``DPG'' and ``KGraph+GD'' are built upon the same approximate \textit{k}-NN graph.
 
As shown from Fig.~\ref{fig:hnsw_vs_app}(a)-(e), the performance trend of KGraph is in general similar as flat-HNSW on synthetic datasets. However it shows considerably worse performance than the rest approaches on the real-world datasets. While KGraph+GD demonstrates stable performance across all the datasets. In particular, it is considerably better than flat-HNSW and HNSW when dimension is greater than \textit{8} on synthetic datasets. When the local intrinsic dimension gets higher, its performance over HNSW becomes more significant. Similar trend is also observed on real-world datasets. The major difference between KGraph+GD, DPG and KGraph lies in the incorporation of graph diversification. As seen from the figure, both GD and DPG boosts the performance of KGraph to the level of HNSW in most of the cases. 

To this end, we find similar performance as HNSW is achievable with flat approximate \textit{k}-NN graph produced in other way around. The advantage of HNSW is only observable when the data dimension is low ($d{\leq}8$). Moreover, the speed efficiency achieved in such case is only two times higher than the one based on flat graphs. Compared to HNSW, flat \textit{k}-NN graph is more attractive in the sense it takes less memory to hold the structure and less time for graph construction. In terms of graph diversification operations, GD is more favorable over DPG for its considerably better performance and the smaller size of the diversified graph than that of DPG. 

As observed from all above experiments, the speed-up that one approach achieves drops steadily as the data intrinsic dimension increases. Graph-based approaches have no exception. On RAND1M and GloVe1M, the speed-ups from all the approaches are very limited.

\section{Curse of dimensionality in Graph-based Search}
\label{sec:exp3}
Based on the above experiments, we learn that either the hierarchical structure in HNSW or graph diversification fails to return descent results on high dimensional data, such as RAND1M. In order to figure out the reason, a close investigation is carried out about HNSW and KGraph+GD as the search is conducted on four synthetic datasets with increasing dimensionality. 

\begin{figure}
	\begin{center}
	\subfigure[RAND10M4D]
		{\includegraphics[width=0.45\linewidth]{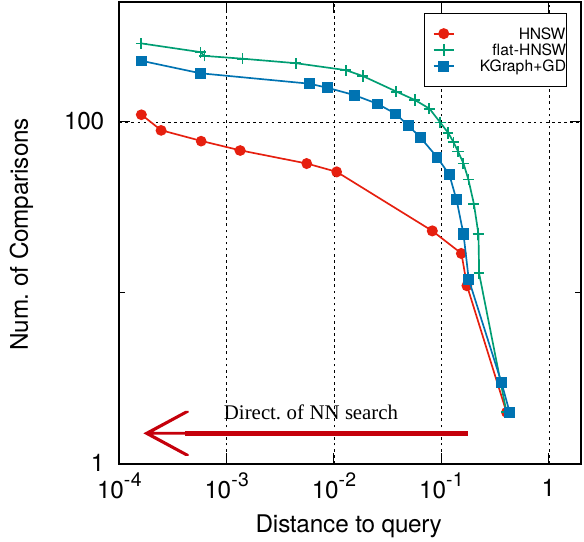}}
	\hspace{0.05in}
	\subfigure[RAND10M16D]
		{\includegraphics[width=0.48\linewidth]{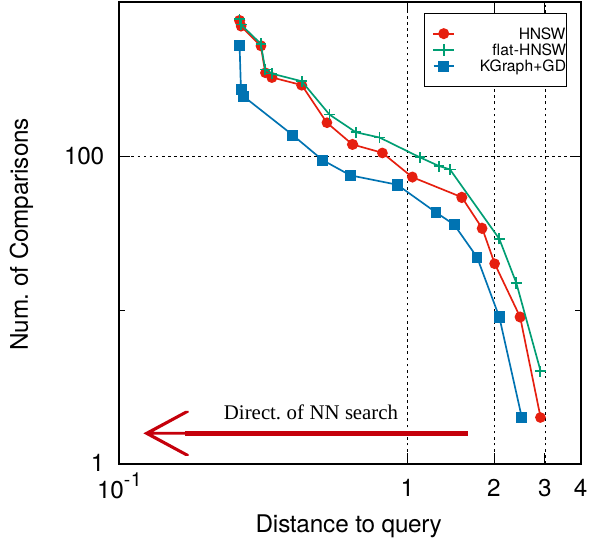}}
		\subfigure[RAND10M32D]
		{\includegraphics[width=0.45\linewidth]{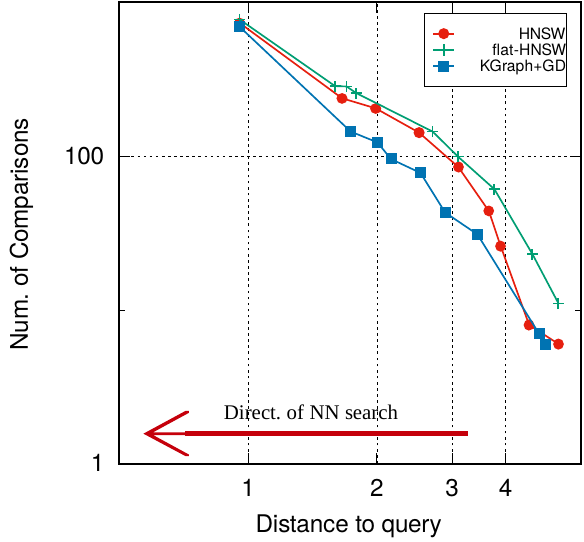}}
	\hspace{0.05in}
	\subfigure[RAND1M100D]
		{\includegraphics[width=0.48\linewidth]{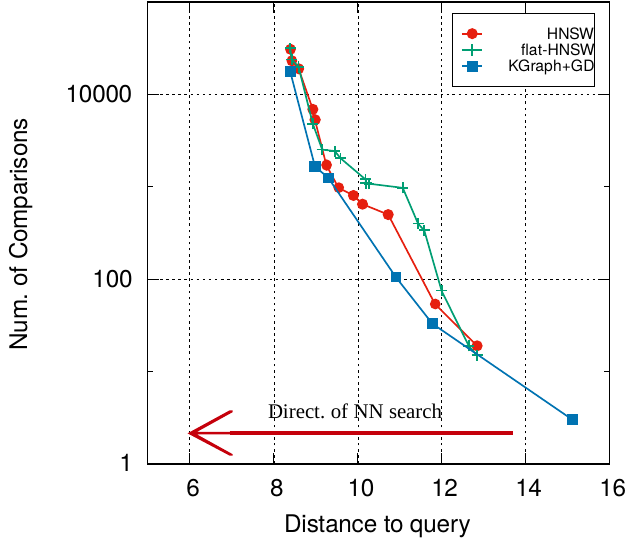}}
	\end{center}
	\caption{The number of comparisons (y-axis) the search spends against the corresponding distance range (\textit{x}-axis) the search reaches to. The observation is conducted on RAND10M4D and RAND1M. The statistics are made with \textit{50} queries for HNSW, flat-HNSW and KGraph+GD on each dataset.}
	\label{fig:qrycost}
\end{figure}

In the experiment, we look into the number of comparisons we need from starting samples (or seeds) to the expected closest neighbor. Specifically, we check the number of comparisons that are spent when the search process reaches to a certain distance range to the query. The statistics are shown in Fig.~\ref{fig:qrycost}.

As shown in the figure, NN search with HNSW does move faster in the early stage of the process than those based on flat graph on both cases. This is more significant on RAND10M4D. As the search advances from distance range $0.5$ to $10^{-2}$, HNSW only takes less than \textit{100} comparisons. While both flat-HNSW and KGraph+GD spend above \textit{100} comparisons. The number of comparisons each approach spends when advancing from distance $10^{-2}$ to $10^{-4}$ is in general similar, which takes up around \textit{15\%} of the total cost. As a result, the cost saved by HNSW on the early search stage becomes significant. However, the situation is largely turned around on the high dimensional case. Although it is observable that HNSW saves up less than \textit{100} comparisons as the search advances from \textit{18.0} to \textit{9.6}, such kind of cost-saving is marginal in contrast to the comparisons spent on ``close neighborhood''. The process spends nearly \textit{10,000} comparisons when it moves from distance \textit{9.6} to around \textit{8.6}. Due to the high cost in ``close neighborhood'' search for all the approaches, there is no big performance difference among three graph based approaches.


\begin{figure}
	\begin{center}
		\includegraphics[width=0.88\linewidth]{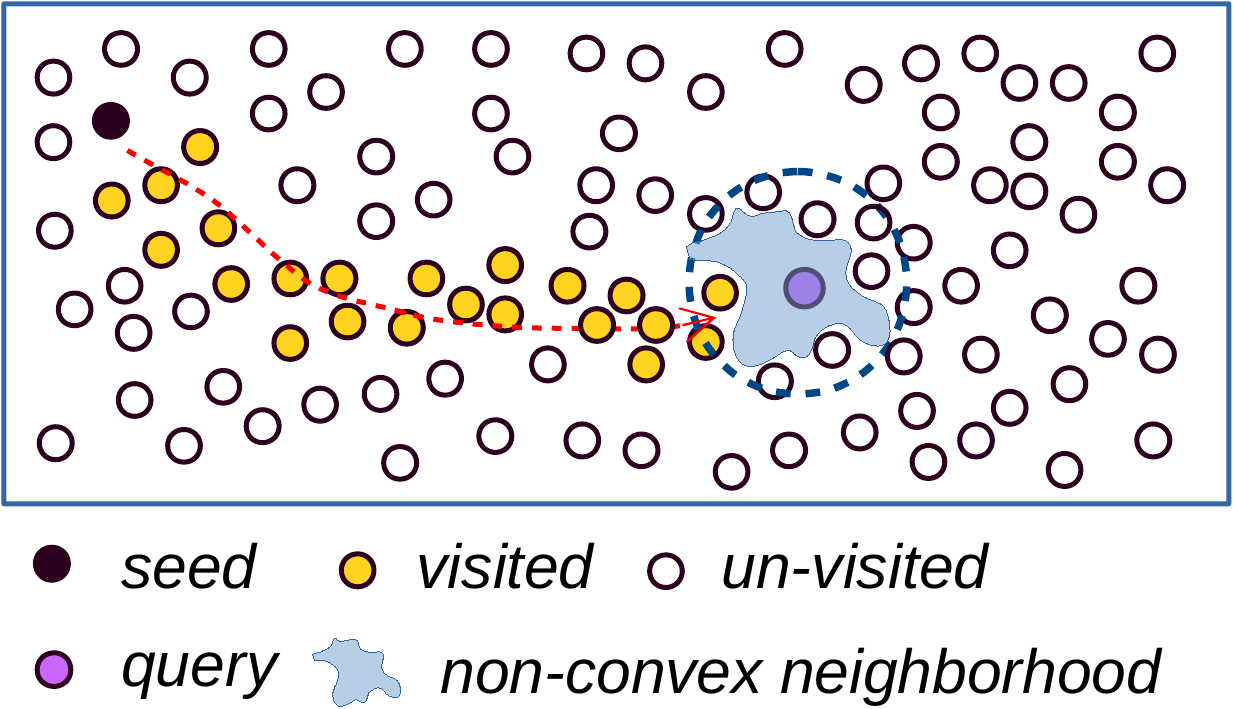}
	\end{center}
	\caption{The illustration of the difficulty faced by hill-climbing and the like search procedures. The search in ``far neighborhood'' colored by red arrow is undertaken largely efficient. While the search in ``close neighborhood'' is inefficient in high dimensional case. It could be as slow as an exhaustive search over the region around.}
	\label{fig:nndescent}
\end{figure}

This phenomenon is interpreted in Fig.~\ref{fig:nndescent}. Please be noted that the ``close neighborhood'' is usually a non-convex region. As the search reaches to the ``close neighborhood'', it could be trapped in a local optima when the maintained top-\textit{k} list is short. Otherwise, when the top-k list is sufficiently long, it might search around the ``close neighborhood'' exhaustively. As a result, it leads to either high efficiency while with low recall or high recall while with low efficiency. In the low dimensional case, this will not be a big problem as samples in this neighborhood reside in the inter-chained NN lists. The hill-climbing therefore converges quickly. In this case, the number of comparisons in ``far neighborhood'' takes a large portion. As a result, the time cost saved by the hierarchical search becomes significant. In the high dimensional case, samples in the ``close neighborhood'' may reside in the NN lists that are isolated from each other. They may be relatively close to each other however hardly overlap with their neighbors. It is therefore hard to guide the search to jump from one local optima to another. As the data dimension goes up higher, this problem becomes more apparent and serious. As a consequence, when both the data dimension and intrinsic data dimension are high, HNSW, as other graph based approaches, becomes inefficient.


\section{Conclusion}
We have presented our comparative studies over two major graph-based approaches. Three major issues are investigated. Firstly, we find that the hierarchical structure of HNSW only shows superior performance over flat graph on low dimensional data. The speed-efficiency over the approaches without hierarchy support is around two times to its best. The advantage that the hierarchical structure brings fades away as the dimension goes up as high as around \textit{32}. Moreover, it is possible to achieve similar or even higher speed efficiency over HNSW with the support of approximate \textit{k}-NN graph, after it has been undergone graph diversification. Furthermore, we find that most of the graph-based approaches face the same difficulty when both data dimension and the intrinsic data dimension are high. This difficulty is directly linked to ``curse of dimensionality''. Neither the hierarchical structure nor the graph diversification scheme, which are effective in many cases, are able to address this issue.

\section*{Acknowledgment}

This work is supported by National Natural Science Foundation of China under grants 61572408. 

\bibliographystyle{ieeetr}
\bibliography{wlzhao}

\vspace{-0.1in}


\end{document}